\documentclass[conference]{IEEEtran}
\IEEEoverridecommandlockouts
\usepackage{cite}
\usepackage{amsmath,amssymb,amsfonts}
\usepackage{algorithmic}
\usepackage{graphicx}
\usepackage{textcomp}
\usepackage{xcolor}
\def\BibTeX{{\rm B\kern-.05em{\sc i\kern-.025em b}\kern-.08em
    T\kern-.1667em\lower.7ex\hbox{E}\kern-.125emX}}
\begin{document}

\title{Avoiding Improper Treatment of Persons with Dementia by Care Robots\\
\thanks{The authors received funding for this work from the Swedish Knowledge Foundation (Sidus AIR no. 20140220 and CAISR 2010/0271), and some travel funding for L. O. from the EU REMIND project (H2020-MSCA-RISE No 734355). We thank all those who kindly helped and contributed comments and thoughts!}
}

\author{\IEEEauthorblockN{Martin Cooney, Sepideh Pashami, Eric J\"{a}rpe, Awais Ashfaq}
\IEEEauthorblockA{\textit{Center for Applied Intelligent Systems Research (CAISR)} \\
\textit{Halmstad University}\\
Halmstad, Sweden \\
martin.cooney, sepideh.pashami, eric.jarpe, awais.ashfaq@hh.se}
\and
\IEEEauthorblockN{Linda Ong}
\IEEEauthorblockA{\textit{I+ srl} \\
Florence, Italy\\
l.ong@i-piu.it}
}

\maketitle

\begin{abstract}
The phrase ``most cruel and revolting crimes'' has been used to describe some poor historical treatment of vulnerable impaired persons by precisely those who should have had the responsibility of protecting and helping them. 
We believe we might be poised to see history repeat itself, as increasingly human-like aware robots become capable of engaging in behavior which we would consider immoral in a human--either unknowingly or deliberately.
In the current paper we focus in particular on exploring some potential dangers affecting persons with dementia (PWD), which could arise from insufficient software or external factors, and describe a proposed solution involving rich causal models and accountability measures:
Specifically, the Consequences of Needs-driven Dementia-compromised Behaviour model (C-NDB) could be adapted to be used with conversation topic detection, causal networks and multi-criteria decision making, alongside reports, audits, and deterrents.
Our aim is that the considerations raised could help inform the design of care robots intended to support well-being in PWD.
\end{abstract}


\section{Promise of dementia care robots}

Interactive robots are expected to be useful for helping to care for and engage persons with dementia (PWD) because dementia is an important problem with high and rising needs which cannot be adequately addressed by the current numbers of human healthcare workers \cite{cooney2018}.
By seeking to slow the progression of dementia, also in PWD who could not previously receive care, and reducing the need for caregivers to travel to remote locations and conduct menial tasks, robots could save money and time.\footnote{Such robots can also be useful by performing other tasks such as cleaning, fetching, and helping in emergencies like when elderly persons fall.}
Moreover, robotic capabilities can be leveraged to facilitate well-being in PWD and caregivers:
\begin{itemize}
\item Machine properties:
\begin{itemize}
\item producability: hardware can be manufactured as required and software knowledge transferred quickly.
\item continuous operation: robots can be active at any time of day or night.
\item imperturbability: robots will not become irritated or bored due to dementia behaviors; e.g., having to answer a question over and over. 
\item communication: robots can be used as a medium to connect people (e.g., through videoconferencing), and can be perceived as easier to confide in than humans.\footnote{For example, an elderly person disclosed a problem of arthritis to a small robot in earshot of caregivers, which could then be addressed \cite{shibata2012paro}} 
\end{itemize}
\item  
Awareness properties:
\begin{itemize}
\item transparency: robots' algorithms and data can be controlled, to seek to avoid human bias.
\item collation: a robot can accurately recall and quickly infer from fragmented big data from various sources (e.g., lifestyle, genetic, patient history), which is difficult and time-consuming for humans \cite{parikh2018}.
\item super-human sensing: robots can sense biosignals such as pulse, temperature, and brainwaves. 
\end{itemize}
\end{itemize}

Based on such motivation, some robots have been designed to be used with PWD. 
Touching small animal-like robots AIBO, NeCoRo, and Paro--which look like a dog, cat, and baby seal--was observed to have various positive effects on persons with mild to severe dementia, such as improved happiness, increased communication, and decreased agitation \cite{yonemitsu2002, libin2004, wada2004}.
Some more complicated interactions have been designed with socially assistive humanoid robots:
Pioneer and Nao robots were programmed to provide cognitive stimulation in therapy sessions related to music, language, storytelling and physiotherapy \cite{tapus2009use, martin2013nao}.
Matilda/PaPeRo and Ryan offer services such as reminders, conversation, and communication; entertainment from dancing, singing and games; as well as information about weather and time; the former has been found in a four year study with over a hundred PWD to have positive emotional and behavioral effects \cite{khosla2017matilda, abdollahi2017ryan}.
And some other robots have been designed to assess cognitive state, such as an ASCC Companion Robot which conducts typical tests like the Mini-Mental State Examination \cite{femandes2017}, and Ludwig which seeks to detect trouble in speech \cite{rudzicz2017ludwig}.
Thus, dementia care robots seem to be becoming more aware, complex, and human-like, in terms of behaviors, emotions, and assessments; although we believe this to be a positive trend in general, some negative consequences could also arise, which should be considered.

\section{Potential dangers}

In addition to common concerns that social robots will replace people or infringe on privacy due to obtrusive sensing, we hypothesize that there is also a danger that history might repeat itself--in the sense that human-like robots might, like humans in the past, sometimes treat persons with impairments improperly.
Specifically, we believe that there have been times when human caregivers had incorrect expectations that their knowledge would be adequate for providing proper treatment, or had to deal with interfering external objectives \cite{Stanley2015}: 
For example, beliefs in witchcraft as a cause for mental disorders, and the efficacy of trepanation, blood-letting, and ice water baths for certain conditions, are no longer considered to be valid. 
And, some practices such as isolating mentally impaired persons were motivated not only by a desire to provide optimal treatment, but also to protect society or reduce a family's burden.
Likewise, robots perceived as having abilities almost at the level of humans could be thrown into roles which they cannot yet sufficiently deal with, and subjected to pressures which erode their capabilities to provide proper treatment.

\subsection{Mistakes}

We expect that care robots could make potentially harmful mistakes due to expectations which are unrealistically high given the complexity of the phenomena involved and the exploratory state of the field.

People's expectations toward human-like robots tend to be too high, and attitudes toward technology can be pragmatic: for example, cars result in many deaths yearly worldwide, but are tolerated due to the convenience they afford.
Moreover, various knowledge is lacking for dementia, behaviors, and emotion: e.g., causes and treatment of Alzheimer's disease, or how to computationally infer the meanings of social signals and represent complex emotions; however, a robot might not be able to always obtain information by directly querying a PWD, who might not remember, know, or be able to sufficiently communicate information.
Additionally, ready-made solutions do not yet exist, and in research ad-hoc solutions are common which might not always take into account  ethical repercussions or the complex confounds which present themselves in human-robot interaction (e.g., Hawthorne effect, demand characteristics, etc.) \cite{wiese2017}.

In particular, our limited knowledge in the area of artificial intelligence has also come under increased scrutiny in recent years, due to identification of some unfair algorithms which unwittingly encode human biases and typically are opaque, damaging, and deployable at massive scales \cite{Oneill2016}.
A troublesome complication is that it can be difficult to hold a robot's creators responsible for harmful algorithms that are not in their exclusive control, according to the principle of ``res ipsa loquitur'' in cyberlaw, as in the case of  the ``Random Darknet Shopper'' which bought drugs on the Dark Web \cite{calo2015}.

\subsection{Deliberate Improper Treatment}

Even with appropriate algorithms, we argue that, like humans, human-like robots could sometimes deliberately provide improper treatment when placed in difficult situations or forced into an abnormal state.
Based on risk factors identified for humans \cite{pillemer1992}, we expect that some potential risks could include if a highly human-like robot must complete crucial tasks on time but is prevented by a PWD--either through violence or disruptive behaviors which cannot be avoided; the robot is placed in a role where it should make decisions for the PWD (no one else can help); the robot has an emotional model which also allows it to feel threatened; and societal norms support improper treatment (e.g., if it is believed that robots are cold and cannot be held accountable for their actions, and that PWD can be treated in an improper manner; this conjecture can be related to the controversial 1971 Stanford Prison Experiment).
A robot's objectives could also conflict in other ways; for example, spending more time with one elderly person can mean less time for another.
Also, trying to keep a PWD healthy could require prohibiting the person from making many of their own choices, as is exemplified in fiction by Nurse Ratched in ``One Flew Over the Cuckoo's Nest'', and the care robot Vera in the televison series ``Real Humans''.

Abnormal behavior has also been alleged to have been conducted by some human caregivers, such as Genene Jones, reportedly the inspiration for the demented nurse in Stephen King's Misery who cripples and intimidates a victim in her care, or Harold Shipman, who is estimated to have killed over 200 victims, mostly elderly women; motives in such cases are not always clear and can range from a desire to achieve financial or work-related goals, to sadism and a desire to dominate or control others.
Some examples of machines which exhibit abnormal behavior have also already been observed: 
Learning systems can be corrupted over time, as in the case of Microsoft's agent ``Tay'' which expressed racist views on Twitter \cite{vincent2016}.
Also, the artificial intelligence (AI)  ``Norman'' was trained to become psychopathic, perceiving horrible events in a Rorschach projective test \cite{wakefield2004}.
We propose that such corruption could occur not only due to data, but also due to polymorphic code, when AIs become capable of adapting their own programs.

This problem will be further compounded if human-like robots become able to lie.
We believe this is an important challenge--also due to current interest in explainable AI, which assumes a system will be truthful--because lying might result in negative repercussions for trust in robots.
The idea of lying robots is not entirely new: In general, systems in which some agents can offer incorrect information are known as ``Byzantine'', and their equilibriatory qualities have been studied via virtual agents in game theory \cite{rehm2005, teacy2006}; also, the possibility of robots to be deceptive has been described, although from the perspective that a human will be responsible for causing a robot to be deceptive \cite{hartzog2014}. 
What is unclear is how to deal with a robot itself choosing to lie in the context of care, or why such a robot might lie.
We believe white lies might be motivated by conflicting requirements.
For example, a robot could untruthfully state that a person's family will visit soon to make a PWD happy, or that certain care is required or has been highly successful in order to promote itself or its organization, such as a care center or pharmaceutical company, for financial gain. 
Black lies could result from robotic psychopathy.

\section{A proposed solution}

We propose (1) the use of rich models informed by human science for identifying causal dynamics, to address the problem of insufficient algorithms, (2) and accountability measures, to guard against deliberate improper treatment.

\subsection{Rich models}

To allow robots to properly care for PWD, we propose leveraging human science models in causal dynamics analysis, emotion-based conversation topic detection, and a conversation strategy also incorporating multi-criteria decision making.

The Consequences of Needs-driven Dementia-compromised Behavior model (C-NDB) describes how surficial treatment of dementia behaviors can fail if underlying needs are not met, also resulting in the emergence of secondary behaviors \cite{kovach2005cndb}.
For example, treating an agitated PWD with antianxiety medicine might be only temporarily effective if the underlying need is arthritic pain requiring painkillers.
Thus, to help a PWD, robots should seek to identify and address such needs.
We propose that causes, needs, and behaviors can be computationally formalized using causal graphs.
However, interventions are required to tailor causal graphs to individuals, because PWD can have very different symptoms.
To detect a PWD's current state, the robot can engage in conversation or receive text input offline (e.g., from a digital diary).
We propose that sentiment analysis can be conducted on a PWD's words to find emotional statements and then conversation topics detected to find problems.
In the simplest case the data could come from a response to a question like ``How are you''?
Conversation-based emotion detection can also be supplemented by detecting other channels such as facial expressions, body heat, or brainwaves.

In the more complicated case of a longer conversation, a conversation strategy can incorporate multi-criteria decision making to balance between accurately determining causes and time taken.
Time taken is not only important if a robot has much work to do, or to avoid badgering with too many questions, but also because it can be difficult to retain the attention of a PWD.
Initially, a robot's questions can be open, and become more specific if a problem is expected; moreover, in the latter case, the robot can seek to vary its queries in the case of absent or repeated responses, and to consider problems of higher urgency first (e.g., following the hierarchies proposed by Maslow \cite{dorenbos2010}).
Moreover, to support good feelings, a robot can seek to match a person's sentiments or to positively distract, although the robot should also not hide the truth, which could decrease a person's trust and acceptance of the robot.
Furthermore, over multiple care sessions, the robot should also conduct change point detection on the causal graphs, which could give insight into disease progression and if problems are being properly addressed. 

\begin{figure}[htbp]
\centerline{\includegraphics[width=0.4\textwidth]{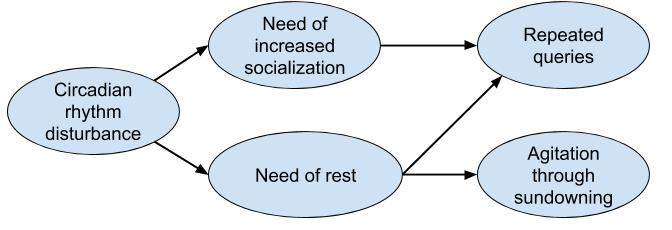}}
\caption{Example of a causal network illustrating the C-NDB Model, in which nodes from left to right represent causes, needs, and behaviors. (Sundowning refers to confusion and restlessness in the evening.)}
\label{fig:causal_graph}
\end{figure}

\begin{figure}[htbp]
\centerline{\includegraphics[width=0.4\textwidth]{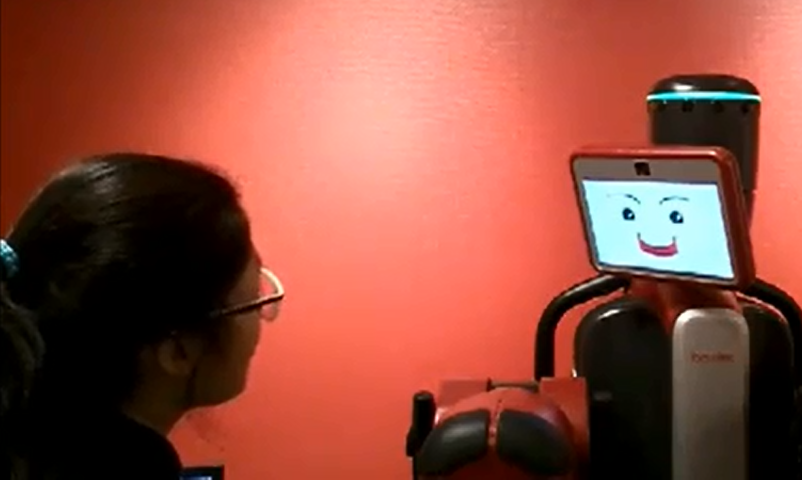}}
\caption{Our robot, Baxter, conducting a proof-of-concept interaction demo in which it seeks to recognize a person's emotional state and respond appropriately. 
Sentiment analysis was conducted using a simplified lexicon-based approach with TextBlob. A video of the interaction is available online at https://www.youtube.com/watch?v=CUoA84k8QxY.}
\label{fig:DB}
\end{figure}

\subsection{Accountability}

To deal with the case in which a robot deliberately provides improper treatment and possibly seeks to conceal its actions, we propose accountability via reports, audits, and deterrents.

\textit{Reports}. In order to realize Byzantine fault tolerance--the capability of a distributed system to function when individual nodes can be unreliable and provide imperfect information--we propose a blockchain-based reporting system for care.
Blockchain simply defined is ``a technology that allows people who don't know each other to trust a shared record of events'' \cite{corea2017}.
For this, post-quantum approaches should be used to ensure security, and interactions could also be summarized, at the cost of potentially missing some important events.\footnote{Even if blockchain is not used, we suggest that reports should be stored in a non-volatile fashion, where they could also serve as a robot's memories, ensuring that some technical glitch does not erase its experiences with humans, and facilitating personalization.}
Additionally, we support the idea of requiring a supplier's declaration of conformity (SDoC) \cite{hind2018}, also for robots: this should furthermore describe the robot's ethics model and any tests conducted; moreover, like the warnings on a tobacco carton or toy box, relevant concerns should be disclosed to both caregivers and PWD.

\textit{Audits}. We propose that interactions should occasionally be monitored by independent systems to reduce the likelihood of improper treatment. For example, continuous monitoring by sensors in a smart environment could indicate problems instantaneously, such as if a PWD becomes injured or cold.
Another possibility is if typical patterns of lying could be detected by other robots;
although detecting lying seems to be highly difficult for humans (even professionals' rates differ little from random chance), a program called ``The Silent Talker'' has been built to classify truth and lying \cite{indvik2009Liar}.

\textit{Deterrents}. A robot's algorithm could be prohibited if it is found to have the potential to cause harm or allow for dishonesty. 
A human-like robot with the capability to be concerned with its own survival could also be encouraged to faithfully serve if its existence is tied to a single PWD; i.e., the robot's software and potentially even hardware could be eliminated when its PWD dies.
To deal with the problem of creators not being persecutable by law for crimes perpetrated by robots, qualities of robots which are deemed to be special in terms of consumer protection law--capability for physical harm, unpredictable action due to learning, and human-likeness--could be kept in check. Then companies using such algorithms could be more easily held responsible and sued or fined.
Limiting learning ability could also help to avoid induced psychopathy like with Tay and Norman, by allowing the robot to adapt to some degree but not to completely invert its value system. Moreover, any changes due to learning should also be reported clearly.

\section{Conclusions}

Thus, rich models informed by human science and accountability measures could be used to mitigate some problems forseen to arise as dementia robots become more advanced, but much remains unclear. For example, although rich personalized models could be used to provide excellent care, ensuring that robots cannot detect individuals, like in blind testing, could potentially better promote fairness. And for accountability, the typical Byzantine assumption is not applicable if all parties deliberately engage in deception--e.g., if a smart environment, caregivers or PWD collude with a robot.
Future work will be required to unravel such tangles to get a better glimpse of the underlying complex conceptual tapestry--toward helping to provide proper care, protect such vulnerable persons who cannot protect themselves, and contribute to their well-being.

\bibliographystyle{IEEEtran}
\bibliography{IEEEabrv,bibliography}


\vspace{12pt}

\end{document}